\documentclass[12pt]{article}
\catcode`\@=11

\@addtoreset{equation}{section}
\def\theequation{\arabic{section}.\arabic{equation}}
     
\catcode`\@=11
\def\thesection{\arabic{section}.}

\def\appendix{\setcounter{section}{0}
        \def\thesection{Appendix.}
        \def\theequation{\Alph{section}.\arabic{equation}}}
\def\section{\@startsection{section}{1}{\z@}{3.5ex plus 1ex minus
   .2ex}{2.3ex plus .2ex}{\large\bf}}
     
\def\eqnarray{\let\@currentlabel=\theequation\refstepcounter{equation}
    \global\@eqnswtrue
    \global\@eqcnt\z@\tabskip\@centering\let\\=\@eqncr
    $$\halign to \displaywidth\bgroup\@eqnsel\hskip\@centering
      $\displaystyle\tabskip\z@{##}$&\global\@eqcnt\@ne 
       \hfil${{}##{}}$\hfil
      &\global\@eqcnt\tw@ $\displaystyle\tabskip\z@{##}$\hfil 
       \tabskip\@centering&\llap{##}\tabskip\z@\cr}
\def\lefteqn#1{\hbox to 4\arraycolsep{$\displaystyle #1$\hss}}
\long\def\@makefntext#1{\parindent 0cm\noindent
\hbox to 1em{\hss$^{\@thefnmark}$}#1}

\newcommand{\beq}{\begin{equation}}
\newcommand{\eeq}{\end{equation}}

\begin{document}

%
%
%
%
\def\citen#1{%
\edef\@tempa{\@ignspaftercomma,#1, \@end, }
\edef\@tempa{\expandafter\@ignendcommas\@tempa\@end}%
\if@filesw \immediate \write \@auxout {\string \citation {\@tempa}}\fi
\@tempcntb\m@ne \let\@h@ld\relax \let\@citea\@empty
\@for \@citeb:=\@tempa\do {\@cmpresscites}%
\@h@ld}
%
\def\@ignspaftercomma#1, {\ifx\@end#1\@empty\else
   #1,\expandafter\@ignspaftercomma\fi}
\def\@ignendcommas,#1,\@end{#1}
%
%
\def\@cmpresscites{%
 \expandafter\let \expandafter\@B@citeB \csname b@\@citeb \endcsname
 \ifx\@B@citeB\relax 
    \@h@ld\@citea\@tempcntb\m@ne{\bf ?}%
    \@warning {Citation `\@citeb ' on page \thepage \space undefined}%
 \else
    \@tempcnta\@tempcntb \advance\@tempcnta\@ne
    \setbox\z@\hbox\bgroup 
    \ifnum\z@<0\@B@citeB \relax
       \egroup \@tempcntb\@B@citeB \relax
       \else \egroup \@tempcntb\m@ne \fi
    \ifnum\@tempcnta=\@tempcntb 
       \ifx\@h@ld\relax 
          \edef \@h@ld{\@citea\@B@citeB}%
       \else 
          \edef\@h@ld{\hbox{--}\penalty\@highpenalty \@B@citeB}%
       \fi
    \else   
       \@h@ld \@citea \@B@citeB \let\@h@ld\relax
 \fi\fi%
 \let\@citea\@citepunct
}
%
\def\@citepunct{,\penalty\@highpenalty\hskip.13em plus.1em minus.1em}%
%
%
\def\@citex[#1]#2{\@cite{\citen{#2}}{#1}}%
%
%
\def\@cite#1#2{\leavevmode\unskip
  \ifnum\lastpenalty=\z@ \penalty\@highpenalty \fi 
  \ [{\multiply\@highpenalty 3 #1
      \if@tempswa,\penalty\@highpenalty\ #2\fi 
    }]\spacefactor\@m}
\let\nocitecount\relax  
%
\begin{titlepage}
\vspace{.5in}
\begin{flushright}
UCD-2001-01\\
gr-qc/0103100\\
January 2001\\
\end{flushright}
\vspace{.5in}
\begin{center}
{\Large\bf
 Liouville Lost, Liouville Regained:\\[1ex] 
 Central Charge in a Dynamical Background}\\
\vspace{.4in}
{S.~C{\sc arlip}\footnote{\it email: carlip@dirac.ucdavis.edu}\\
       {\small\it Department of Physics}\\
       {\small\it University of California}\\
       {\small\it Davis, CA 95616}\\{\small\it USA}}
\end{center}

\vspace{.5in}
\begin{center}
{\large\bf Abstract}
\end{center}
\begin{center}
\begin{minipage}{4.75in}
{\small
Several recent approaches to black hole entropy obtain the density of states
from the central charge of a Liouville theory.  If Liouville theory is coupled 
to a dynamical spacetime background, however, the classical central charge 
vanishes.  I show that the central charge can be restored by introducing
appropriate constraints, which may be interpreted as fall-off conditions
at a boundary such as a black hole horizon.
}
\end{minipage}
\end{center}
\end{titlepage}
\addtocounter{footnote}{-1}

In an effort to understand the microscopic degrees of freedom responsible for
black hole entropy, a number of researchers have recently begun to look at
Liouville theories, either at spatial infinity \cite{Coussaert,Navarro,Bautier,%
Bautier2,Rooman,Skenderis,Krasnov} or in a neighborhood of the horizon 
\cite{Solodukhin}.  Such theories naturally arise at the asymptotic boundary
of the (2+1)-dimensional BTZ black hole \cite{Coussaert}, and are relevant to
many higher-dimensional black holes whose near-horizon behavior resembles 
that of the BTZ black hole \cite{Strom}; they may also be obtained near the 
horizon of an arbitrary black hole by dimensionally reducing to the $r$--$t$
plane \cite{Solodukhin}.  Since Liouville theory is a conformal field theory,
its density of states can be inferred from its central charge by means of the 
Cardy formula \cite{Cardy}.  There is some debate as to whether the Liouville 
states represent the genuine gravitational degrees of freedom or merely give an 
effective  ``thermodynamic'' description \cite{Krasnov,Martinec,Frolov,Carlip1}, 
but in either case, the result offers a potential explanation for the universality of
the Bekenstein-Hawking entropy: the density of states may be determined by
conformal symmetry, independent of the details of quantum gravity \cite{Carlip2}.

The computation of black hole entropy from Liouville theory depends sensitively
on the central charge.  In particular, it is the appearance of a {\em classical\/}
central charge---that is, a central term that is already present in the Poisson
brackets---that leads to an order $1/\hbar$ contribution to the entropy.  
Unfortunately, as I shall demonstrate below, when Liouville theory is coupled 
to a dynamical two-dimensional metric, the classical central charge vanishes.
This problem is evaded in some approaches to Liouville theory at spatial infinity,
in which the boundary conditions freeze the metric, but it is present in other
treatments of spatial infinity \cite{Bautier,Bautier2,Rooman,Skenderis} and in
the near-horizon derivation \cite{Solodukhin}.

Of course, one can recover the central charge, and thus the Bekenstein-Hawking 
entropy, by freezing the dynamics of the metric.  But this seems too strong a
condition.  The main goal of this paper is to demonstrate that it is sufficient to 
impose asymptotic fall-off conditions on the metric, either at infinity or at the 
horizon.  To show this, I will introduce a new method for treating such fall-off 
conditions, via second class constraints and Dirac brackets, which I hope may be 
more generally useful.

\section{Liouville central charge in a dynamic background}

We begin with a generalized Liouville action,
\beq
I_L[\varphi, g] = {1\over4\pi}\int d^2x \sqrt{-g}\left\{
   {1\over2}g^{ab}\partial_a\varphi\partial_b\varphi 
   + {1\over\gamma}\varphi R + V[\varphi]\right\} .
\label{a1}
\eeq
For standard Liouville theory, the potential is 
\beq
V[\varphi] = {\mu\over2\gamma^2}e^{\gamma\varphi} ,
\label{a2}
\eeq
but more general forms are possible \cite{Seiberg}.  After the usual ADM
decomposition of the metric,  
\beq
ds^2 = N^2dt^2 - \sigma^2\left(dx + \beta dt\right)^2 ,
\label{a3}
\eeq
a tedious but straightforward computation brings the action to the canonical 
form
\beq
I_L = \int dt \int dx \, \left( 
  \pi_\sigma{\dot\sigma} + \pi_\varphi{\dot\varphi} 
   - N{\cal H} - \beta{\cal P} \right) ,
\label{a4}
\eeq
with canonical momenta
\begin{eqnarray}
\pi_\sigma &=& 
    {1\over 2\pi\gamma}{1\over N}\left({\dot\varphi} - \beta\varphi'\right) ,
    \nonumber\\
\pi_\varphi &=& 
    {1\over 4\pi}{\sigma\over N}\left({\dot\varphi} - \beta\varphi'\right) 
    + {1\over 2\pi\gamma}{1\over N}\left({\dot\sigma} - (\beta\sigma)'\right)
\label{a5}
\end{eqnarray}
and Hamiltonian and momentum constraints
\begin{eqnarray}
{\cal H} &=& 2\pi\gamma \pi_\varphi\pi_\sigma 
   - {\pi\over 2}\gamma^2 \sigma\pi_\sigma^2
   - {1\over2\pi\gamma}\left({\varphi'\over\sigma}\right)' 
   + {1\over8\pi}{\varphi^{\prime 2}\over\sigma}
   - {1\over4\pi}\sigma V[\varphi]  , \nonumber\\
{\cal P} &=& \varphi'\pi_\varphi - \sigma\pi_\sigma' .
\label{a6}
\end{eqnarray}
(Primes denote derivatives with respect to $x$, dots derivatives 
with respect to $t$.) 

The symmetries of the canonical theory are the ``surface deformations,'' 
generated by
\beq
H[\hat\xi] = \int dx\,{\hat\xi}{\cal H} , \quad
P[\hat\eta] = \int dx\,{\hat\eta}{\cal P} .
\label{a7}
\eeq
These are closely related to diffeomorphisms: given a vector field 
$(\xi^t,\xi^x)$, the transformation 
\beq
\delta F = \left\{H[N\xi^t] + P[\xi^x+\beta\xi^t], F\right\}
\label{a8}
\eeq
of any function on phase space is equal on shell to the Lie derivative 
${\cal L}_\xi F$.  The surface deformation group splits into two chiral 
pieces, generated by
\beq
H^\pm[{\hat\xi}] =  {1\over2}\left\{
    H[{\sigma\hat\xi}] \pm P[{\hat\xi}]\right\}  .
\label{a9}
\eeq
Another long but straightforward computation then shows that
\begin{eqnarray}
\left\{ H^\pm[{\hat\xi}],H^\pm[{\hat\eta}]\right\} &=& 
    \pm H^\pm [({\hat\xi}{\hat\eta}' - {\hat\eta}{\hat\xi}')]
    \nonumber\\
\left\{ H^+[{\hat\xi}],H^-[{\hat\eta}]\right\} &=& 0 .
\label{a10}
\end{eqnarray}

As expected, we have obtained two chiral copies of the Virasoro algebra.
But perhaps unexpectedly, instead of the Liouville central charge 
$c = 12/\gamma^2$, there is no central term in the algebra.  Quantization
may introduce a central charge, of course, but as noted in the introduction,
the quantum central charge has the wrong dependence on $\hbar$ to be
useful to explain black hole entropy.

One way to understand what has happened is to write the dynamical metric
$g_{ab}$ as
\beq
g_{ab} = e^{\gamma\chi}{\bar g}_{ab} ,
\label{a11}
\eeq
where ${\bar g}_{ab}$ is a constant curvature metric, determined by a finite
number of moduli.  Then using the standard Weyl transformation properties
of the scalar curvature,
\beq
R[g] = e^{-\gamma\chi}\left( R[{\bar g}] - \gamma{\bar\Delta}\chi \right) ,
\label{a12}
\eeq
we find that
\beq
I_L[\varphi, g] = I_L[\varphi + \chi, {\bar g}] - I_L[\chi,{\bar g}] .
\label{a13}
\eeq
The dynamical metric thus induces a Liouville action with the opposite sign,
and the opposite central charge, of the original action, leading to a cancellation
of the central term.  This is again a known result, although now ``read backwards.''  
It has long been understood that any conformal theory of ``matter'' coupled to
a Liouville theory of ``gravity'' has vanishing central charge \cite{DDK}; here,
our ``matter'' is itself a Liouville field, while gravity was treated canonically,
but essentially the same arguments apply.
 
\section{Freezing the metric with Dirac constraints}

It is clear from eqn.\ (\ref{a13}) that we should be able to recover the standard
central charge by freezing the metric $g_{ab}$ (and thus the field $\chi$).  
This result can, of course, be obtained directly from the action (\ref{a1}) by 
fixing the metric \cite{Teitelboim}.  But for later purposes, it is useful to instead 
fix the metric by means of constraints in the Poisson algebra.

As our first constraint, we fix the spatial metric $\sigma$:
\beq
C_1(x) = \sigma(x) - \sigma_0(x) = 0
\label{b1}
\eeq
where $\sigma_0$ is a fixed spatial metric, that is, a (not necessarily constant) 
metric that has vanishing Poisson brackets with all phase space variables.  As 
our second constraint, we demand that $\sigma_0$ be time-independent; from 
eqn.\ (\ref{a5}), this means that
\beq
C_2(x) = \pi_\varphi(x) - {\gamma\over2}\sigma\pi_\sigma(x) = 0 .
\label{b2}
\eeq

Note that $C_1$ and $C_2$ do not commute.  Indeed, from the canonical Poisson
brackets,
\beq
\{C_1(x), C_2(y) \} =  -{\gamma\over2}\sigma\delta(x-y) .
\label{b3}
\eeq
In Dirac's terminology,  $C_1$ and $C_2$ are second class constraints.  Such 
constraints modify the Poisson brackets \cite{Dirac,HennTeit}: one must construct 
a new bracket $\{\ ,\, \}^*$ to preserve the constraints.  Specifically, let $K_{ij}(x,y)$ 
be the kernel
\beq
\int dy\, \sum_j K_{ij}(x,y)\{C_j(y),C_k(z)\} = \delta_{ik}\delta(x-z) ,
\label{b4}
\eeq
where $\{\ , \}$ is the ordinary Poisson bracket.  The Dirac bracket is
then
\beq
\{A(x),B(y)\}^* = \{A(x),B(y)\} - \int dz_1 \int dz_2 \sum_{i,j}
   \{A(x),C_i(z_1)\}K_{ij}(z_1,z_2)\{C_j(z_2),B(y)\} ,
\label{b5}
\eeq
and $\{A(x),C_i(y)\}^*$ is identically zero.

For our Liouville theory, it follows from (\ref{b3}) that
\beq
K_{12}(z_1,z_2) = {2\over\gamma}\sigma^{-1}\delta(z_1-z_2) ,
\label{b6}
\eeq
and it is easy to verify that
\begin{eqnarray}
\left\{ H^\pm[{\hat\xi}], C_1 \right\} 
    &=& \mp{1\over2}\left(\sigma\xi\right)' \\
\left\{ H^\pm[{\hat\xi}], C_2 \right\} 
    &=& -{1\over4\pi\gamma}\left( \xi' + {\sigma'\over\sigma}\xi \right)' 
    - {1\over8\pi}\xi\sigma^2\left( {dV\over d\varphi} - \gamma V\right) .
\nonumber
\label{b7}
\end{eqnarray}
If $V[\varphi]$ is the Liouville potential (\ref{a2}), the last term in the
bracket of $H^\pm$ with $C_2$ vanishes.  The Dirac brackets of the
original constraints then become
\begin{eqnarray}
\left\{ H^\pm[{\hat\xi}],H^\pm[{\hat\eta}]\right\}^* &\approx& 
    \pm H^\pm [({\hat\xi}{\hat\eta}' - {\hat\eta}{\hat\xi}')] \nonumber\\
    &&\pm {1\over4\pi\gamma^2}\int dz \left[
    \left( \xi' + {\sigma'\over\sigma}\xi \right)
          \left( \eta' + {\sigma'\over\sigma}\eta \right)' -
     \left( \eta' + {\sigma'\over\sigma}\eta \right)
          \left( \xi' + {\sigma'\over\sigma}\xi \right)' 
     \right]  
    \nonumber\\
\left\{ H^+[{\hat\xi}],H^-[{\hat\eta}]\right\}^* &\approx& 0 ,
\label{b8}
\end{eqnarray}
where, following Dirac, ``weak equality'' ($\approx$) means ``equality up to
terms proportional to the constraints.''  If we now shift $H^\pm[{\hat\xi}]$ 
by appropriate functions of $\sigma$,
\beq
{\tilde H}^\pm[{\hat\xi}] = H^\pm[{\hat\xi}] +
    {1\over4\pi\gamma^2}\int dz \xi \left[ \left({\sigma'\over\sigma}\right)^2
    - 2\left({\sigma'\over\sigma}\right)' \right]
\label{b9}
\eeq
(note that $\{H^\pm[\xi],\sigma\}^*\approx0$), we find
\begin{eqnarray}
\left\{ {\tilde H}^\pm[{\hat\xi}],{\tilde H}^\pm[{\hat\eta}]\right\}^* &\approx& 
    \pm {\tilde H}^\pm [({\hat\xi}{\hat\eta}' - {\hat\eta}{\hat\xi}')] 
    \pm {1\over4\pi\gamma^2}\int dz \left[ \xi'\eta'' -  \eta'\xi'' \right]  
    \nonumber\\
\left\{ {\tilde H}^+[{\hat\xi}],{\tilde H}^-[{\hat\eta}]\right\}^* &\approx& 0 ,
\label{b10}
\end{eqnarray}
describing two commuting Virasoro algebras with classical central charges
\beq
c^\pm = {12\over\gamma^2} ,
\label{b11}
\eeq
thus recovering the standard Liouville result \cite{Seiberg}.

\section{Central charge from near-boundary conditions}

While the method used in the preceding section is new, the conclusion,
of course, is not.  But the method gains power when one realizes that
the constraints (\ref{b1})--(\ref{b2}) need not be applied globally:
it is enough to impose them as fall-off conditions near a boundary.

To see this, it is useful to reinterpret the Dirac brackets (\ref{b5}) in a
manner suggested by Bergmann and Komar \cite{Bergmann}.  Consider
a system with second class constraints $\{C_i\}$, and let $A$ be a
function on phase space that is a candidate for a physical observable.  
The constraints vanish on the space of physically admissible fields, so 
if $A$ has nonzero Poisson brackets with any of the $C_i$, it cannot 
really be an observable.  But we can always shift $A$ by  a term proportional 
to the constraints without altering the physics.  In particular, if we set
\beq
A^* = A + \sum_i \lambda_iC_i\quad \hbox{with}\ \lambda_i =
    - \sum_j\{A,C_j\}K_{ji} 
\label{c1}
\eeq
with $K_{ij}$ defined as in eqn.\ (\ref{b4}), it is evident that $\{A^*,C_k\}
\approx0$.  If we now consider two functions $A$ and $B$, it is easy to
show that the Dirac bracket of $A$ and $B$ is simply the ordinary
Poisson bracket of $A^*$ and $B^*$:
\beq
\{A,B\}^* = \{A^*,B^*\} .
\label{c2}
\eeq

Now let us return to our Liouville theory, and suppose that we have a boundary
at $x=x_0$.  Let us try to impose the constraints (\ref{b1})--(\ref{b2}) only
in a small neighborhood $[x_0,x_0+\epsilon]$ of the boundary.  The standard
Dirac procedure now fails, since $\{C_1,C_2\}$ is no longer invertible, but the
Bergmann-Komar variation extends trivially: we need merely add a suitable
multiple of the constraints in the region $[x_0,x_0+\epsilon]$.  The Poisson
brackets (\ref{b10}) then become
\begin{eqnarray}
\left\{ {\tilde H}^\pm[{\hat\xi}],{\tilde H}^\pm[{\hat\eta}]\right\}^* &\approx& 
    \pm {\tilde H}^\pm [({\hat\xi}{\hat\eta}' - {\hat\eta}{\hat\xi}')] 
    \pm {1\over4\pi\gamma^2}\int_{x_0}^{x_0+\epsilon} dz 
         \left[ \xi'\eta'' -  \eta'\xi'' \right]  
    \nonumber\\
\left\{ {\tilde H}^+[{\hat\xi}],{\tilde H}^-[{\hat\eta}]\right\}^* &\approx& 0 ,
\label{c3}
\end{eqnarray}
where the only change is that the region of integration in the central term is now 
restricted to a neighborhood of the boundary.  The brackets (\ref{c3}) are still
those of a pair of commuting Virasoro algebras with central charge (\ref{b11}),
now defined on a strip.  In particular, the central charge is independent of the
distance $\epsilon$: the constraints need be imposed only in an arbitrarily small
neighborhood of the boundary.

It would be worthwhile to develop a slightly different approach, in which the
constraints (\ref{b1})--(\ref{b2}) are imposed as fall-off conditions near a
boundary: that is, we require $C_1\sim0$ and $C_2\sim0$ near $x=x_0$,
where ``$\sim 0$'' means ``approaches zero sufficiently rapidly.''  I will not
work out details here, but if we choose a coordinate $x$ that is independent 
of phase space (and thus has vanishing Poisson brackets with all fields), it is
clear that the Bergmann-Komar approach can be adapted to conditions 
$C_i = O((x-x_0)^n)$ to determine an asymptotic algebra of constraints.

\section{Future steps: black holes and dilaton gravity}

We have now seen that even in a dynamical spacetime background, the
central charge of a Liouville theory can be recovered through the imposition
of suitable boundary conditions.  This means  that Solodukhin's ``near-%
horizon conformal field'' approach to back hole entropy \cite{Solodukhin}
is consistent after all.  

The results above also suggest a new approach to determining $L_0$, the 
value of the Virasoro zero mode needed to apply Cardy's formula to the 
counting of states.  A starting point is eqn.\ (\ref{b9}), which tells us that 
when the near-horizon metric is nonconstant, the Virasoro generators
must be shifted.  For conformal coordinates near a black hole horizon,
\beq
{\sigma'\over\sigma} = \kappa = {2\pi\over\beta} ,
\label{d1}
\eeq
where $\kappa$ is the surface gravity and $\beta$ is the inverse Hawking 
temperature.  If we look at diffeomorphisms with period $L$,
\beq
\xi_n = {L\over2\pi}e^{2\pi inx/L}
\label{d2}
\eeq
(the normalization is fixed by the condition that $\{ \xi_m,\xi_n\} =
\xi_{m+n}$), eqn.\ (\ref{b9}) yields
\beq
L_0 = {1\over2\gamma^2}\left({L\over\beta}\right)^2 ,
\label{d3}
\eeq
and thus by Cardy's formula,
\beq
S = 2\pi\sqrt{cL_0\over6} = {2\pi\over\gamma^2}{L\over\beta} .
\label{d4}
\eeq

Now, in Solodukhin's analysis of dimensionally reduced spherically
symmetric gravity in $d$ dimensions\footnote{Note that Solodukhin's 
normalization of the Liouville action differs from the one used here.} 
\cite{Solodukhin},
\beq
{1\over\gamma^2} = q^2 \left({d-3\over d-2}\right) 
    {A_{\hbox{\scriptsize hor}}\over8G} ,
\label{d5}
\eeq
where $q$ is an undetermined parameter coming from a field redefinition.
This gives the right general form for the Bekenstein-Hawking entropy.
If we further demand that
\beq
c = {3A_{\hbox{\scriptsize hor}}\over2\pi G} ,
\label{d6}
\eeq
as determined by the boundary analysis of Ref.\ \cite{Carlip3}, then
(\ref{d4}) yields an entropy
\beq
S = {A_{\hbox{\scriptsize hor}}\over4G}{L\over\beta} ,
\label{d7}
\eeq
giving the correct Bekenstein-Hawking entropy for the natural choice of
periodicity $L=\beta$.

This is clearly not a complete explanation of black hole entropy, since
it requires a condition (\ref{d6}) that has not been obtained within the
Liouville formalism, but it might be more than a mere coincidence.  To 
investigate this issue further, it may be useful to look at a similar analysis
of dilaton gravity with more explicitly geometric boundary conditions
at the horizon.  Work on this problem is in progress.

\vspace{1.5ex}
\begin{flushleft}
\large\bf Acknowledgements
\end{flushleft}

This work was supported in part by Department of Energy grant
DE-FG03-91ER40674.

\end{document}